# Keldysh theory re-examined: Application of the generalized Bessel functions

Jarosław H. Bauer[*]

*Katedra Fizyki Teoretycznej, Wydział Fizyki i Informatyki Stosowanej Uniwersytetu Łódzkiego, Ul. Pomorska 149/153, PL-90-236 Łódź, Poland*

A derivation of the ionization rate for the hydrogen-like ion in the strong linearly polarized laser field is presented. This derivation utilizes the famous Keldysh probability amplitude in the length gauge (in the dipole approximation) and without Coulomb effects in the final state of the ionized electron. No further approximations are being made, because the amplitude has been expanded in the double Fourier series in a time domain (with the help of the generalized Bessel functions). Thus, our theory has no other limitations characteristic of the original Keldysh theory. We compare our "exact" theory with the original Keldysh one, studying photoionization energy spectra and total ionization rates. We show breakdown of the original Keldysh theory for higher frequencies. In the barrier-suppresion regime the "exact" Keldysh theory gives results closer to well-known numerical or other analytical results.

___________________________________

[*]bauer@uni.lodz.pl

## I. INTRODUCTION

One of the most important and the most frequently used methods employed to study an ionization in a strong laser field are $S$-matrix theories [1-3]. They are usually called Keldysh, or Keldysh-Faisal-Reiss (KFR) theories [4-7]. Also the name "strong-field approximation" (SFA) is sometimes preferred [7]. One usually makes the assumption that an ionized electron, due to an action of the laser field, makes a transition from a given initial bound state to a final free state, which can be treated as dominated only by the laser field. In light of the presented considerations the simplest version of the theory neglects completely an effect of a binding potential (short-range or long-range) and uses the Gordon-Volkov wave function [8] as a final state of the ionized electron. In his pioneering work [4] Keldysh identified the so-called adiabaticity parameter $\gamma$ (termed after him) and obtained relatively simple analytical expressions for ionization rates and arbitrary $\gamma$. The Keldysh parameter $\gamma$ is often used to distinguish between multiphoton ionization (when $\gamma \gg 1$) and tunneling ionization (when $\gamma \ll 1$).

$$\gamma = \frac{\omega\sqrt{2E_B}}{F} = \frac{Z\omega}{nF} , \qquad (1)$$

where $\omega$ is the laser frequency, $F$ - the amplitude of the laser field, and $E_B = Z^2/(2n^2)$ - the binding energy of the atom (of the nuclear charge $Z$) initially in the state described by the well-known $(n,l,m)$ quantum numbers (without spin), in the nonrelativistic approach. In Ref. [4] and in the present paper we consider only the hydrogen-like ion (i.e. the initial state is the ground state: $(n=1,0,0)$) in the linearly polarized monochromatic laser field. Such a simple theory can be applied only when the laser field is strong enough. An extensive and a very good discussion of applicability conditions of the SFA can be found in Refs. [1,7] (see also Sect. III in Ref. [9]). Based on this, we restrict our interest here only to the domain of the field parameters $(\omega, F)$, where $\gamma < 1$ (preferably when $\omega \ll F$) and the motion of the ionized electron remains nonrelativistic. Currently Eq. (21) from Ref. [4] which describes ionization rate for $\gamma > 1$ has only a character of an estimation and a more accurate theory should take into account Coulomb effects in the final state of the ionized electron. As it was noted by Keldysh and others [10,11], there are two conditions that are necessary to derive final expressions for the ionization rate. These conditions limit the applicability range of the theory: (i) the saddle point approximation to perform the contour integral, and (ii) the small final kinetic momentum approximation. According to Ref. [10], the adiabatic assumption $\omega << E_B$ is necessary, to fulfill the condition (i). There have been attempts to derive ionization rate formulas (starting from Eq.(2)) without building on (i) or (ii) [10-12]. However, we are aware of no expressions involving the generalized Bessel functions in the length gauge (for linear polarization). In this way our present work fills some gap in the literature. On the other hand, the saddle-point method or semi-analytical methods enabled much progress in the last few decades (see, for example, Refs. [2,3,13-16]).

In contrast, the well-known expressions derived in the velocity gauge for $H^-$ ion and the $H(1s)$ atom long ago by Reiss [7] include Bessel functions (the ordinary ones for circular polarization and the generalized ones for linear polarization). These formulas do not rely on the assumptions (i) and (ii). The main aim of the present work is to derive expressions analogous to those in Ref. [7] in the length gauge and investigate their main properties. In the case of circularly polarized laser field such theory has been developed for arbitrary initial states $(n,l,m)$ with $n=1$ and $n=2$ of the hydrogen-like ion [9,17,18]. $S$-matrix theories in the above-mentioned two gauges give qualitatively similar results (particularly this concerns the shape of photoelectron energy spectra) for initial states of even parity [19], like the ground states of $H(1s)$ and $H^-$. However, several earlier calculations have convinced us that the length gauge gives ionization rates closer to experimental results, and theoretical static-field results (when $\omega << E_B$) [9,20,21] (and references therein). It seems that in the so-called barrier-suppression regime relevance of Coulomb corrections in the final state of the ionized electron should not be great and should decrease with increasing the laser field $F$. Thus it appears reasonable to study the case without Coulomb corrections in details. This could be a kind of some benchmark result and a starting point for future improvements (including Coulomb corrections).

Our paper is organized as follows. In Sec. II we present the basis of the original Keldysh theory [4] and its main results for $\gamma < 1$. In Sec. III we present our analytical calculations. The main result is given in Eq. (16). In Sec. IV we discuss the results of our numerical calculations (including photoelectron energy spectra and total ionization rates) and we make a comparison of these results with other results, mainly with the original Keldysh theory. In the Appendix we show in details how the Fourier coefficients and the generalized Bessel functions have been calculated in this paper. In the present work we consistently use atomic units (a.u.): $\hbar = e = m_e = 1$, substituting explicitly -1 for the electronic charge. We keep any nuclear charge $Z$ in all the equations given below, but finally, in our numerical calculations, we put $Z = 1$ for the hydrogen atom.

## II. KELDYSH THEORY

In the Keldysh theory [4] one starts from the following matrix element, which is the approximate probability amplitude of strong-field ionization in the length gauge:

$$(S-1)_{fi}^{Keldysh} = -i \int_{-\infty}^{\infty} dt \int d^3r \, \Psi^{GV}(\vec{r},t)^* \vec{r} \vec{F}(t) \Phi_i(\vec{r},t) \,, \qquad (2)$$

where the initial, ground state of a hydrogen-like ion is described by the well-known wave function

$$\Phi_i(\vec{r},t) \equiv \Phi_i(\vec{r}) \exp(iE_B t) = \sqrt{\frac{Z^3}{\pi}} \exp(-Zr) \exp\left(i\frac{Z^2}{2}t\right) \qquad (3)$$

in position space. In the length gauge the laser-atom interaction Hamiltonian is $H_I = \vec{r}\vec{F}(t)$, with $\vec{F}(t) = -\frac{1}{c}\frac{\partial \vec{A}}{\partial t}$ being the electric field vector, and the Gordon-Volkov state [8] in this gauge is

$$\Psi^{GV}(\vec{r},t) = \frac{1}{(2\pi)^{3/2}} \exp\left[ i\vec{\pi}(t)\vec{r} - \frac{i}{2}\int_{-\infty}^{t}\vec{\pi}(t)^2 d\tau \right], \qquad (4)$$

where $\vec{p}$ is kinetic (asymptotic) momentum of the ionized outgoing electron and its canonical momentum is given by

$$\vec{\pi}(t) = \vec{p} + \frac{1}{c}\vec{A}(t). \qquad (5)$$

In the dipole approximation magnetic-field component of the laser is zero and the electric-field one is $\vec{F}(t) = \vec{\varepsilon} F \sin(\omega t + \varphi_0)$ (where $\vec{\varepsilon}$ is the polarization vector and $\varphi_0$ - some arbitrary initial phase of the laser field). Starting from Eq. (1), Keldysh [Eq. (20) in Ref. [4]] obtained the following ionization rate formula (for $\gamma < 1$):

$$W^{Keldysh} \approx \frac{\sqrt{3\pi ZF}}{2^{7/4}} \exp\left[ -\frac{2Z^3}{3F}\left(1 - \frac{1}{10}\gamma^2\right)\right]. \qquad (6)$$

In the above expression $n$-photon contributions have been summed up owing to the procedure valid when $\gamma \to 0$ (going over from summation over $n$ to integration). Without this procedure, the ionization rate (6) has the general form

$$W^{Keldysh} = \sum_{n=0}^{\infty} W_n(E_B, \omega, F), \qquad (7)$$

where $W_n$ are partial ionization rates corresponding to absorption of exactly $n$ photons above threshold. $W_n$ are given by Eqs. (16), (18), and (19) in Ref. [4]. To obtain Eq. (6) one also has to make a Taylor expansion around $\gamma = 0$, leaving only the lowest-order term in the pre-exponential factor and the two lowest-order terms in the exponent of Eq. (16) in Ref. [4]. Now then, one can also write down "an intermediate" expression, which does not involve "a multiphoton" summation, but contains all the complicated dependence on the Keldysh parameter $\gamma$:

$$W^{Keldysh} \approx \sqrt{\frac{\sqrt{2}}{\omega}} \sqrt{2E_B} \omega \left(\frac{\gamma}{\sqrt{1+\gamma^2}}\right)^{3/2} \frac{\sqrt{3\pi}}{4\gamma^2} \exp\left[\frac{-E_B(2+\gamma^{-2})}{\omega}\left(\ln\left(\gamma+\sqrt{1+\gamma^2}\right) - \frac{\gamma\sqrt{1+\gamma^2}}{1+2\gamma^2}\right)\right]$$
(8) .

We have left all the above factors not simplified to show the form similar to Eq. (16) in Ref. [4]. However, after careful verification, we have found that the latter equation has a small error, namely the factor $\sqrt{\sqrt{2}/\omega}$ is missing in Ref. [4]. Equation (8) is compatible with Eq. (6) when $\gamma \to 0$. Equation (8) behaves much better than Eq. (6) for $\gamma \approx 1$ (even for $\gamma > 1$ the function (8) is monotonic near $\gamma = 1$, unlike the function (6), cf. Figs. 8 and 9). Of course, when $\gamma << 1$ Eqs.(6) and (8) give nearly identical results. We have also verified numerically that for $F = const$ and $\omega \to 0$ Eq. (8) goes to finite value (like Eq. (6)), as should be. To obtain the most accurate expression (7), one has to include the missing factor $\sqrt{\sqrt{2}/\omega}$ as well. Let us denote these three ionization rates as "Keldysh 1" (Eq. (7)), the most accurate), "Keldysh 2" (Eq. (8)), and "Keldysh 3" (Eq. (6), the least accurate), respectively.

**III. KELDYSH THEORY WITH THE GENERALIZED BESSEL FUNCTIONS**

Using Eqs. (3)-(5) one can show [9] that the ionization probability amplitude (2) may be presented as follows

$$(S-1)^{Keldysh}_{fi} = i \int_{-\infty}^{\infty} dt \tilde{\Phi}_i(\vec{\pi}(t))\left(\frac{1}{2}\vec{\pi}(t)^2 + E_B\right)\exp\left[\frac{i}{2}\int_{-\infty}^{t}\vec{\pi}(\tau)^2 d\tau + iE_B t\right],$$
(9)

where the momentum-space wave function of the initial ground state of the $H(1s)$ atom is equal to

$$\tilde{\Phi}_i(\vec{p}) = \int \frac{d^3 r}{(2\pi)^{3/2}} \exp(-i\vec{p}\vec{r})\Phi_i(\vec{r}) = \frac{\sqrt{8Z^5}}{\pi} \frac{1}{(Z^2 + p^2)^2} .$$
(10)

To calculate the ionization probability amplitude (9) we assume that the laser field propagates along the $x$ axis and the polarization vector $\vec{\varepsilon}$ is parallel to the $z$ axis. Following Reiss [1,7] (and our earlier works, for example [9]) we introduce the $z$ parameter such that $U_P = z\omega = I/(4\omega^2)$, where $U_P$ denotes the ponderomotive energy of the electron and $I = F^2$ denotes the laser intensity in atomic units ($1\ a.u. = 3.51 \cdot 10^{16} W/cm^2$). The product of the first two factors in the integrand of Eq. (9) is proportional to

$$\left(\frac{1}{2}\vec{\pi}(t)^2 + E_B\right)^{-1} = \sum_{k=-\infty}^{\infty} A_k(\vec{p}) e^{ik(\omega t + \varphi_0)}, \tag{11}$$

and may be expanded in the above Fourier series. It appears that the coefficients can be calculated analytically with the help of the residue theorem (see the Appendix). The exponential factor in Eq. (9) may be also expanded, in the standard way, using the Fourier-Bessel expansion:

$$\exp[ia\sin(\omega t + \varphi_0) + ib\sin 2(\omega t + \varphi_0)] = \sum_{n=-\infty}^{\infty} J_n(a,b) \exp[in(\omega t + \varphi_0)], \tag{12}$$

where $a$ and $b$ do not depend on time, and $J_n(a,b)$ are generalized Bessel functions of two arguments, which appear as a result of the integral $\int^t \vec{\pi}(\tau)^2 d\tau$ for linearly polarized field (we use the same convention with respect to $J_n(a,b)$ as in Ref. [7]). When we apply expansions (11) and (12), the amplitude (9) turns out to be proportional to

$$(S-1)_{fi}^{Keldysh} \sim i \sum_{n,k=-\infty}^{\infty} J_n(a,b) A_k(\vec{p}) \exp[i(n+k)\varphi_0] \int_{-\infty}^{\infty} dt \exp\left[i\left(\frac{p^2}{2} + z\omega + (n+k)\omega + E_B\right)t\right], \tag{13}$$

where $a = 2\sqrt{z/\omega}\, p\cos\vartheta$, $b = z/2$, and $\vartheta$ is the angle between $\vec{p}$ and $\vec{\varepsilon}$. Let us note that the summation over $n$ can be replaced by the summation over $N = n+k$. In this way the amplitude is proportional to

$$(S-1)_{fi}^{Keldysh} \sim i \sum_{N,k=-\infty}^{\infty} J_{N-k}(a,b) A_k(\vec{p}) \exp[iN\varphi_0] \int_{-\infty}^{\infty} dt \exp\left[i\left(\frac{p^2}{2} + z\omega + N\omega + E_B\right)t\right], \quad (14)$$

The integral upon time leads to the well-known distribution (a particular model of Dirac $\delta$ function), which enables applying standard procedure. The differential ionization rate $w(\vec{p})$, which is the transition probability per unit time and unit volume in the canonical momentum ($\vec{p}$) space, can be found from

$$w(\vec{p}) = \lim_{t\to\infty} \frac{|(S-1)_{fi}|^2}{t} . \quad (15)$$

To obtain the total ionization probability per unit time $W$, one has to integrate the differential ionization rate over all the possible final momenta of the outgoing electron. The final result is

$$W = \int w(\vec{p}) d^3p = \sum_{N=N_0}^{\infty} W_N = \sum_{N=N_0}^{\infty} Z^5 \sqrt{8E_N} \int_0^{\pi} d\vartheta \sin\vartheta \left(\sum_{k=-\infty}^{\infty} (-1)^k A_k(E_N,\vartheta) J_{N+k}(a,b)\right)^2 ,$$

$$(16)$$

where $a = \sqrt{8zE_N/\omega}\cos\vartheta$ and $b = z/2$. The minimal number of photons absorbed is $N_0 = [z + E_B/\omega] + 1$, and the kinetic energy of the ionized outgoing electron is

$$E_N = p_N^2/2 = N\omega - z\omega - E_B . \quad (17)$$

The symbol $[..]$ denotes integer part of the (positive) number inside. In deriving equation (16) we have used the following property of the generalized Bessel functions:

$J_{-n}(a,b) = (-1)^n J_n(a,-b)$, and the relation: $A_{-k}(\vec{p}) = A_k(\vec{p})$ (see the Appendix for some other details).

Let us note that analogous expression in the velocity gauge is much simpler [7] and contains only one summation. In our notation it is given below ( $a$ and $b$ as in Eq.(16)):

$$W^{SFA} = \int w^{SFA}(\vec{p}) d^3 p = \sum_{N=N_0}^{\infty} \frac{Z^5 \sqrt{8E_N}}{(E_N + E_B)^2} \int_0^\pi d\vartheta \sin\vartheta J_N(a,b)^2 \ . \qquad (18)$$

Equations (16) and (18) show total ionization rates as a sum over partial $N$-photon ionization rates, where the kinetic energy of outgoing electron is given by Eq. (17). The same concerns Eq. (7), but there the index $n$ denotes the number of photons above threshold. Thus $n = N - N_0$. This is obvious, if we look at the argument of the Dirac $\delta$ function from Eq. (14) in Ref. [4].

## IV. NUMERICAL RESULTS AND CONCLUSION

The original Keldysh theory [4] has a high-frequency limitation. In our case ($Z=1$) the theory is expected to be valid, if $\omega < E_B = Z^2/2 = 0.5$ *a.u.* (preferably when $\omega << E_B$). In Figs. 1-7 we present several photoionization energy spectra of outgoing electrons for laser frequencies obeying this condition and the condition $\gamma < 1$. In each of these plots we show three curves, corresponding, respectively, to the most accurate original "Keldysh 1" [Eq.(7)] spectrum, to present "Exact Keldysh" [Eq.(16)] and to its velocity gauge counterpart "SFA (Reiss)" [Eq.(18) and Ref. [7]]. In Figs. 1-3 $\omega = 0.01$ *a.u.* and the peak laser field $F$ decreases from Fig. 1 to Fig. 3. Comparing upper two curves each time, one can see that agreement between "Keldysh 1" and "Exact Keldysh" is quite satisfactory, particularly for lower photoelectron energies. This is understandable, because the small final kinetic momentum approximation is utilized in "Keldysh 1". According to this approximation only electrons with sufficiently small final kinetic energy $E$ should mostly contribute to the total ionization rate ($E = p^2/2 \ll E_B = 0.5$ *a.u.*). This is hardly satisfied only in Fig. 3. In Figs. 1, 2 and 4-7 the range of significant kinetic energies is much wider. With increasing $F$ more and more electrons have $E \geq \sim 0.5$ *a.u.* (see Figs. 2 and 1). This concerns all three curves, but velocity gauge partial ionization rates are at least one order of magnitude (or more) smaller than length gauge ones. Let us note that oscillations that appear in the "Exact Keldysh" and "SFA (Reiss)" curves are not of constant period, which is much greater than photon energy. This shows some affinity between these two results, which have been obtained with the help of the generalized Bessel functions. In Figs. 4-7 we keep the Keldysh parameter constant: $\gamma = 0.33 \ll 1$, which is quite typical in many experiments. In Figs. 4-7 we increase the laser frequency from $\omega = 0.02$ *a.u.*, through 0.057 *a.u.*, 0.1 *a.u.* up to 0.25 *a.u.*, respectively. With increasing the laser frequency (and the peak laser field) the "Keldysh 1" curves become more and more different from respective "Exact Keldysh" curves. This is particularly visible for higher energies, where small momentum approximation fails. For $\omega = 0.1$ *a.u.* and $\omega = 0.25$ *a.u.* "Keldysh 1" theory overestimates the number of high-energy electrons in a significant way. It is worth to recall that the SFA has no frequency limitation [1,7] (except the fact that $\omega$ should not be in resonance with excited bound states of the ionized atom), so its length gauge counterpart derived here ("Exact Keldysh") does not have this limitation as well.

In Figs. 8 and 9 we show various total ionization rates as a function of the peak laser field in the range of two orders of magnitude: $0.02$ *a.u.* $\leq F \leq 2$ *a.u.* (this corresponds to four orders of magnitude in intensity) for $\omega = 0.1$ *a.u.* and $\omega = 0.2$ *a.u.*, respectively. Thus, the barrier-suppression ionization range is included here. Several years ago Bauer and Mulser in Ref. [22] concluded that no analytical theory is able to properly describe the ionization rate in this range for $\omega \ll E_B$. They proposed a simple numerical fit (based on *ab initio* calculations) for the $H(1s)$ atom, namely $W = 2.4F^2$. This line is shown in our log-log plots as a slanted dashed line. Figures 8 and 9 resemble Fig. 6 from Ref. [22]. (However, let us notice that numerical ionization rates for $\omega = 0.1$ *a.u.* and $\omega = 0.2$ *a.u.* in the latter figure indicate that the proper ionization rate should be a concave function in log-log plot, but not a ruled line.) It seems that a significant theoretical progress was attained about ten years later [14]. In Figs. 8 and 9 for comparison we show also the ionization rate of Popruzhenko *et al.* (see Eqs. (5)-(7) from Ref. [14]), which is valid for any Keldysh parameter and takes into account Coulomb effects in the final state of the ionized electron. Curiously enough, our "Exact Keldysh" curve typically lies about one order of magnitude below the curve denoted as "Ref. [14]" and the distance between these two curves decreases with increasing $F$. In Figs. 8 and 9 we show also for comparison the above-mentioned three "Keldysh 1, Keldysh 2, Keldysh 3" ionization rates (marked as "K1, K2, K3" in brief) as well. These three curves converge for $\gamma \to 0$ (as should be) and (except for the region where $F \geq \sim 1\ a.u.$) lie below our "Exact Keldysh" curve. In Figs. 8 and 9 we also give two vertical dashed lines corresponding to $\gamma = 1$ and $\gamma = 0.1$. (The curves "K1, K2, K3" are partially shown also for $\gamma > 1$ only in order to better distinguish them.)

In conclusion, we have derived the ionization rate formula for the hydrogen-like ion in the strong linearly polarized laser field, using the Keldysh probability amplitude in the length gauge (in the dipole approximation) and without Coulomb effects in the final state of the ionized electron. This calculation is exact in the sense that no further analytical approximations are used. As one could expect, it appears that the original Keldysh theory [4] (where additional analytical approximations have been done) leads to satisfactory photoelectron energy spectra only in its low-energy part and only when $\omega << E_B$. The total ionization rate is affected moderately by these approximations, which cause an underestimation of this rate by a factor of 2-3 at best. The price to be paid for the lack of additional analytical approximations are numerical calculations connected with the generalized Bessel functions. These calculations become really time-consuming for low frequencies or large Keldysh parameters for the method applied by us (see the Appendix for more details). Another price which we pay is connected with analytical calculations of the Fourier coefficients from Eq. (11). However, since it is possible for the $H(1s)$ atom, it is likely possible for other bound states of this atom. There have been many attempts to take into account Coulomb effects in the final state of the ionized electron (see, for example, Refs. [9,20] and references therein) by replacing the Gordon-Volkov wave function ($\Psi^{GV}$) by various more complicated Coulomb-Volkov wave functions ($\Psi^{CV}$). These latter are approximate solutions of the full time-dependent Schrödinger equation. It is obvious, that the general scheme of calculations presented here could be readily repeated, if $\Psi^{CV}$ is used instead of $\Psi^{GV}$. This would only change the Fourier coefficients $A_k(\vec{p})$ (cf. Eq. (11)). Our investigations in this direction are under way.

## ACKNOWLEDGMENTS

The present paper has been supported by the University of Łódź.

## APPENDIX

Let us consider the Fourier expansion of the expression $\left(\vec{\pi}(t)^2/2 + E_B\right)^{-1}$ as a function of time. Multiplying both sides of Eq. (11) by $\exp[-in(\omega t + \varphi_0)]$, and integrating this equation from $0$ to $T = 2\pi/\omega$ we obtain

$$A_n(\vec{p}) = \frac{\omega}{2\pi} \int_0^T \frac{\exp[-in(\omega t + \varphi_0)]\, dt}{\left(\vec{p} + \vec{A}(t)/c\right)^2/2 + E_B} \,, \tag{A1}$$

$A_n(\vec{p})$ is always a real number. Let us introduce new parameters: $A = p^2/2 + z\omega + E_B > 0$, $B = 2\sqrt{z\omega}\, p\cos\vartheta$, and $C = z\omega > 0$. Then Eq. (A1) takes the form

$$A_n(\vec{p}) = \frac{\omega}{2\pi} \int_0^T \frac{\exp[-in(\omega t + \varphi_0)]\, dt}{A + B\cos(\omega t + \varphi_0) + C\cos 2(\omega t + \varphi_0)} \,. \tag{A2}$$

The denominator in Eq. (A2) is always positive (for real times $t$), so when going to complex time domain we obtain poles beyond real axis. Substituting the new complex variable $s = \exp[-i(\omega t + \varphi_0)]$, the integral (A2) may be transformed into

$$A_n(\vec{p}) = \frac{1}{\pi i} \oint_{C^+} \frac{s^{n+1} ds}{Cs^4 + Bs^3 + 2As^2 + Bs + C} , \tag{A3}$$

where $C^+$ denotes a circle of a unit radius, with the center in the origin, in the complex s plane. The "plus" means that we circulate the circle in a counter-clockwise direction. To utilize the residue theorem, we have to solve the equation:

$$Cs^4 + Bs^3 + 2As^2 + Bs + C = 0 . \tag{A4}$$

We note that the above equation has symmetric coefficients. In such a case, one can solve Eq. (A4) by the substitution $u = s + 1/s$, obtaining a quadratic equation for $u$, and finally a quadratic equation for $s$. In this way one obtains analytically the four different complex roots of the Eq. (A4), namely $s_1$, $s_2$, $s_3$, and $s_4$. The roots are functions of $A$, $B$, and $C$. As a result, the roots are quite complicated functions of $p$, $\vartheta$, $\omega$, $z$, and $E_B$, so we do not give these complicated expressions here. In practice (in our numerical calculations) we were solving two quadratic equations by turns in a complex domain. The roots $s_1$, $s_2$, $s_3$, and $s_4$ may be numbered so as to $s_2 = s_1^*$, $s_3 = 1/s_1$, $s_4 = s_3^*$, and $|s_1| = |s_2| < 1$ (the asterisk denotes complex conjugate). Then $|s_3| = |s_4| > 1$ and only $s_1$ and $s_2$ contribute to the sum over residues. The final expression is

$$A_n(\vec{p}) = \frac{4}{C} \mathrm{Re} \frac{s_1^{n+1}}{(s_1 - s_2)(s_1 - s_3)(s_1 - s_4)} . \tag{A5}$$

The above Fourier coefficients and the generalized Bessel function obey some sum rules, for example, for any real numbers $a$ and $b$ (see [7,23])

$$\sum_{n=-\infty}^{\infty} J_n(a,b) = 1 , \qquad \text{and} \qquad \sum_{n=-\infty}^{\infty} J_n^2(a,b) = 1 . \tag{A6}$$

Let us consider any bound-state momentum-space wave function $\tilde{\Phi}_{bound}(\vec{p})$ in any binding potential. We assume that $\tilde{\Phi}_{bound}(\vec{p})$ is normalized to unity in the entire space. Hence, the following equation is satisfied for any instant of time

$$\int d^3p \left|\tilde{\Phi}_{bound}(\vec{\pi}(t))\right|^2 = \int d^3p \left|\tilde{\Phi}_{bound}\left(\vec{p} + \frac{1}{c}\vec{A}(t)\right)\right|^2 = 1 . \tag{A7}$$

Making the Fourier expansion of $\tilde{\Phi}_{bound}(\vec{p} + \vec{A}(t)/c)$, like in Eq. (11), substituting it to Eq. (A7), and integrating both sides of the resultant equation over time from 0 to $T = 2\pi/\omega$ we obtain

$$\sum_{n=-\infty}^{\infty} \int d^3p \left|F_n(\vec{p})\right|^2 = 1 , \tag{A8}$$

where $F_n(\vec{p})$ denote the respective Fourier coefficients. If $\tilde{\Phi}_{bound}(\vec{p})$ is the bound state in the zero-range potential (the Dirac $\delta$ one, with $E_B = Z^2/2$) then (see, for example, [9])

$$\tilde{\Phi}_{bound}(\vec{p}) = \frac{\sqrt{Z}}{2\pi} \frac{1}{p^2/2 + E_B} . \tag{A9}$$

Thus, $\tilde{\Phi}_{bound}(\vec{p} + \vec{A}(t)/c)$ is simply equal to the left-hand side of Eq. (11) times $\sqrt{Z}/(2\pi)$. It follows then from Eq. (A8) that

$$\sum_{n=-\infty}^{\infty} \int d^3p \left|A_n(\vec{p})\right|^2 = \frac{4\pi^2}{Z} . \tag{A10}$$

In numerical calculations connected with Eq. (16) one has to cut somewhere infinite summations over $k$ and $N$. The range of these summations should not be too narrow. Consecutive terms $W_N$ in Eq. (16) are positive and usually decrease with increasing $N$ (see Figs. 1-7), so there is no problem with finding the sum with a given relative accuracy. Consecutive Fourier coefficients $A_k(\vec{p})$ decrease with increasing $|k|$, and Eq. (A10) helps us to find the respective range of the values of $k$. We have also tested our numerical subroutines for computing $J_n(a,b)$. They obey the sum rules (A6) with the relative accuracy much lower than 1. In the present work the difficulty connected with finding the ionization rate analytically (with the help of saddle-point method or residue theorem) is essentially replaced by the difficulty connected with finding numerically the generalized Bessel functions $J_n(a,b)$. There is no problem with numerical calculations for low values of $n$, $a$, and $b$. However, for strong laser fields, when $\gamma < 1$ (particularly for $\gamma \ll 1$ or $\omega \ll 1$), arguments and orders of these Bessel functions become large, and one has to proceed in a specific way [7,23]. One can apply a set of asymptotic formulas for $J_n(a,b)$, depending on relative magnitudes of $n$, $a$, and $b$ [24,25] (and references therein). One can also utilize recurrence relations for $J_n(a,b)$, but this should be done with care to avoid problems with numerical stability [26]. In this work we have chosen yet another method to compute $J_n(a,b)$, namely a direct numerical integration of the equation defining the generalized Bessel functions:

$$J_n(a,b) = \frac{1}{\pi}\int_0^\pi d\phi \cos(a\sin\phi + b\sin 2\phi - n\phi) . \tag{A11}$$

When either of the numbers: $n$, $a$, $b$ is large, the integrand is highly oscillatory and passes zero many times in the interval $[0,\pi]$. Especially in this case the so-called linear approximation model is very useful [27]. Let us consider the following integral

$$I = \int_\alpha^\beta dx \cos[kf(x)] , \tag{A12}$$

where $f(x)$ is a smooth function, $|f(x)| \le 1$, and $|k| \gg 1$. The interval $[\alpha,\beta]$ can be divided into $J \gg 1$ equal subintervals, in such a way that $x_j = \alpha + (2j-1)\delta$ denotes the center of the $j$-th subinterval ($j = 1,2,...,J$, and $\delta = (\beta-\alpha)/(2j)$). Then the integral $I$ is equal to

$$I = \sum_{j=1}^{J} \int_{x_j-\delta}^{x_j+\delta} dx \cos[kf(x)] \ . \tag{A13}$$

In the $j$-th subinterval one can substitute a new variable $t = kf(x) \approx kf(x_j) + kf'(x_j)(x-x_j)$, where a Taylor expansion has been applied. This approximation is justified for sufficiently small $\delta$, because $|x-x_j| \leq \delta$ and $|f'(x)|$, $|f''(x)|,\ldots$ (and so on) are typically of the same order of magnitude as $|f(x)|$ (cf. Eq. (A11)). Since the function $t(x)$ is linear, regardless of $f(x)$, one can easily find $I$ analytically:

$$I = \sum_{j=1}^{J} 2\cos[kf(x_j)] \frac{\sin[kf'(x_j)\delta]}{kf'(x_j)} \ . \tag{A14}$$

The above expression behaves well when $f'(x_j)=0$. The fraction in Eq. (A14) is then simply equal to $\delta$. We were also using the efficient Gaussian procedure [28] to claculate numerically the integral from Eq. (A11) and to verify the results obtained from Eq. (A14) in an independent way.

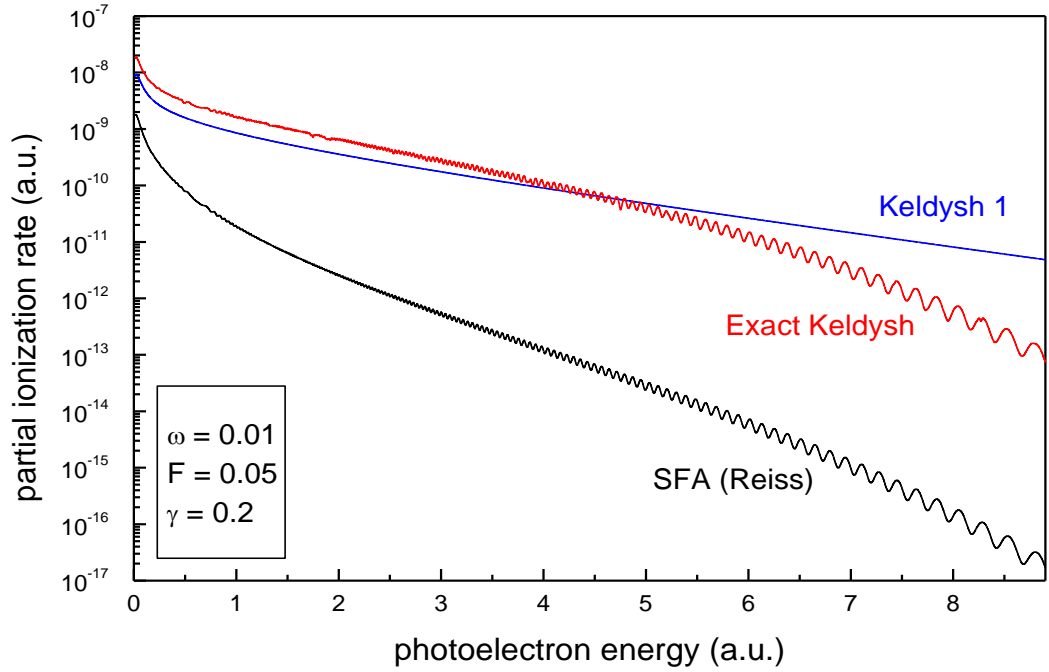

FIG. 1. (Color online) Photoelectron energy spectra for "Keldysh 1" theory [Eq. (7) and Ref. [4]] (solid blue line), "Exact Keldysh" theory [Eq. (16)] (solid red line) and "SFA (Reiss)" theory [Eq. (18) and Ref. [7]] (solid black line). Laser field parameters and the Keldysh parameter are given in the text frame of the plot. (See the main text for more details.)

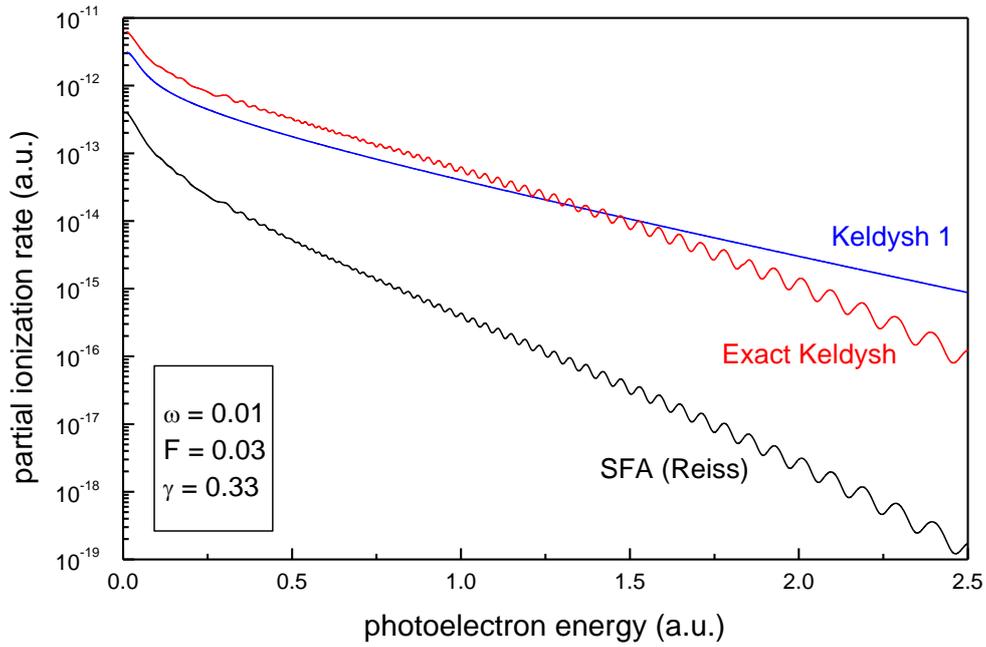

FIG. 2. (Color online) Same as Fig. 1, but for $\omega = 0.01$ *a.u.*, $F = 0.03$ *a.u.*, and $\gamma = 0.33$.

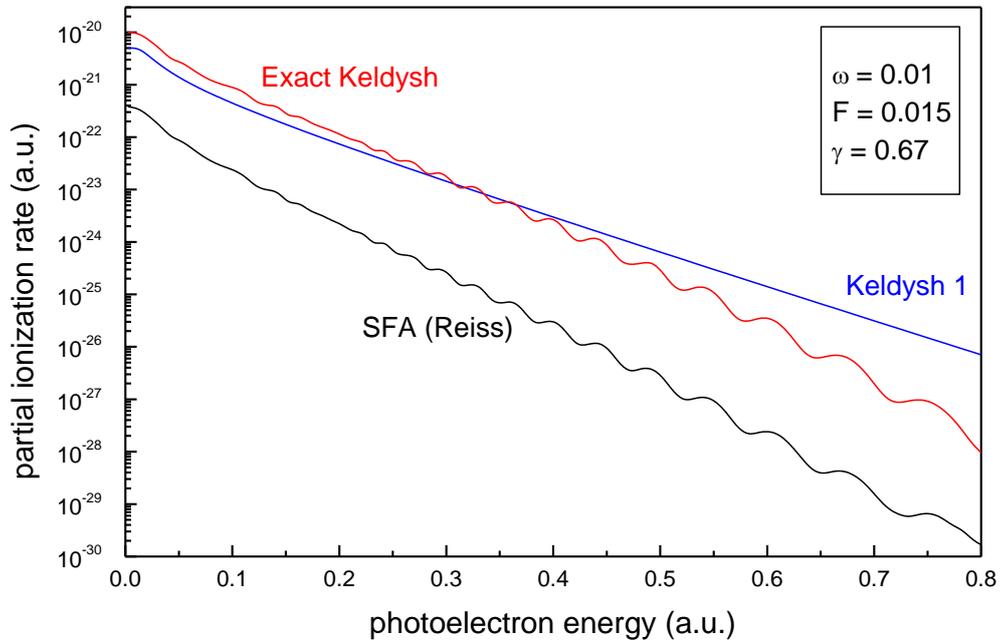

FIG. 3. (Color online) Same as Fig. 1, but for $\omega = 0.01$ *a.u.*, $F = 0.015$ *a.u.*, and $\gamma = 0.67$.

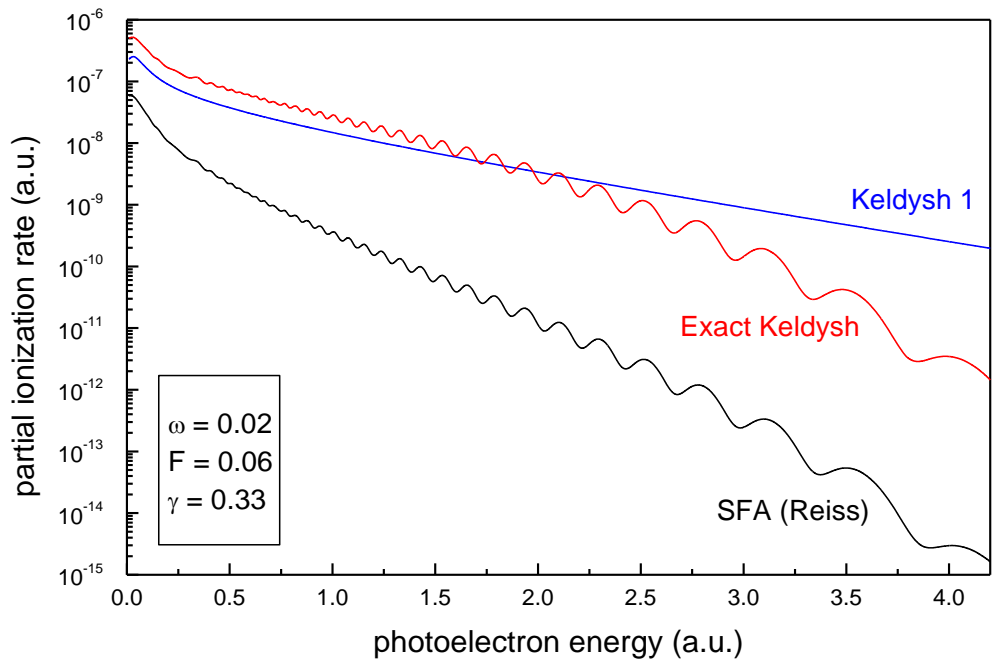

FIG. 4. (Color online) Same as Fig. 1, but for $\omega = 0.02$ *a.u.*, $F = 0.06$ *a.u.*, and $\gamma = 0.33$.

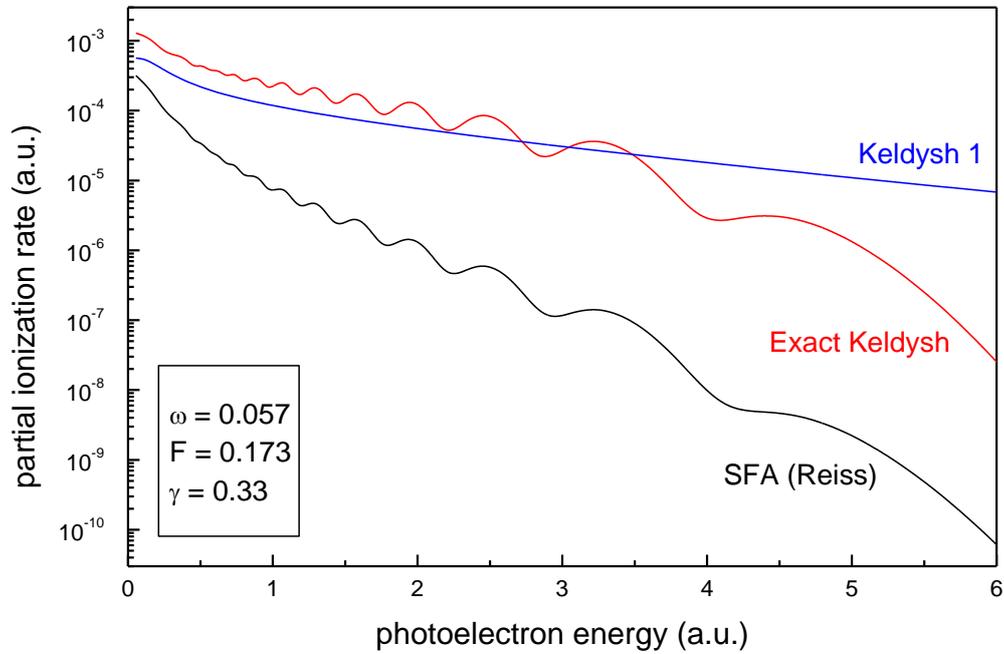

FIG. 5. (Color online) Same as Fig. 1, but for $\omega = 0.057$ *a.u.*, $F = 0.173$ *a.u.*, and $\gamma = 0.33$.

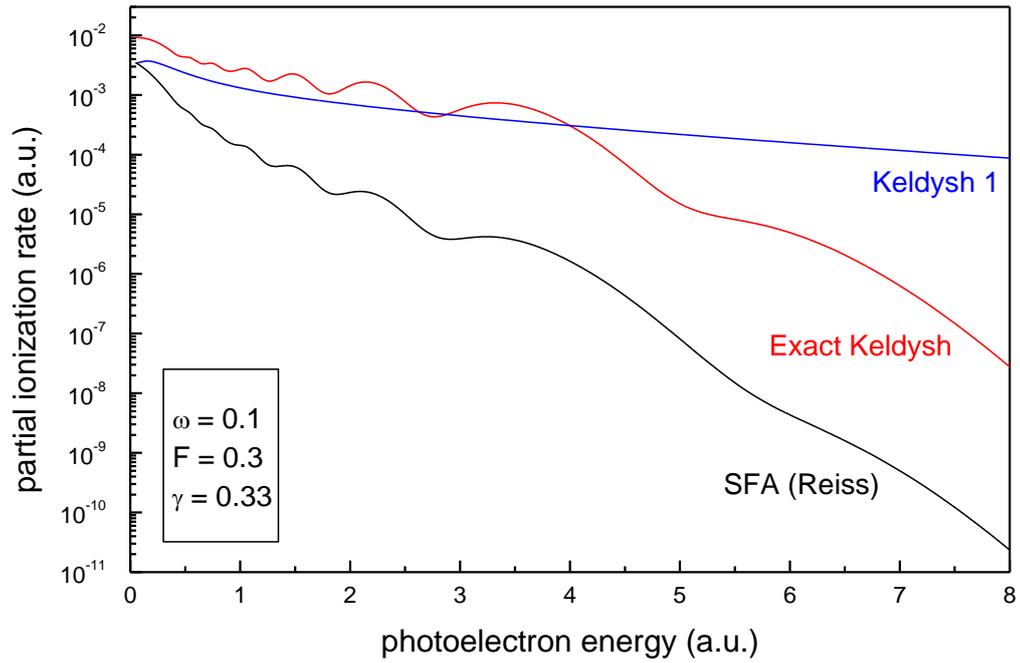

FIG. 6. (Color online) Same as Fig. 1, but for $\omega = 0.1$ $a.u.$, $F = 0.3$ $a.u.$, and $\gamma = 0.33$.

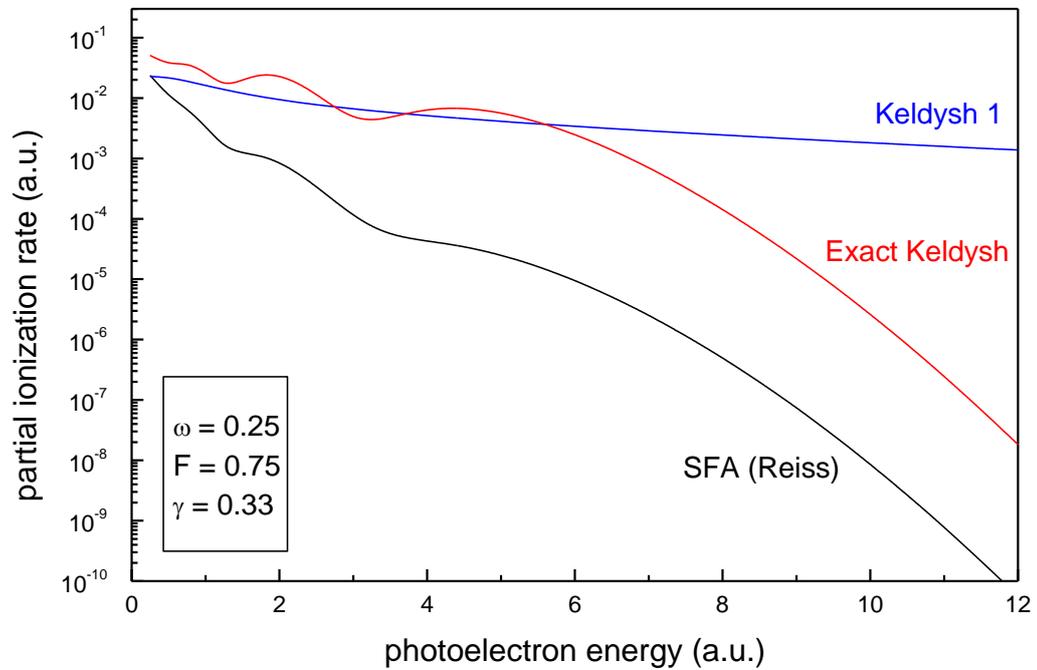

FIG. 7. (Color online) Same as Fig. 1, but for $\omega = 0.25$ $a.u.$, $F = 0.75$ $a.u.$, and $\gamma = 0.33$.

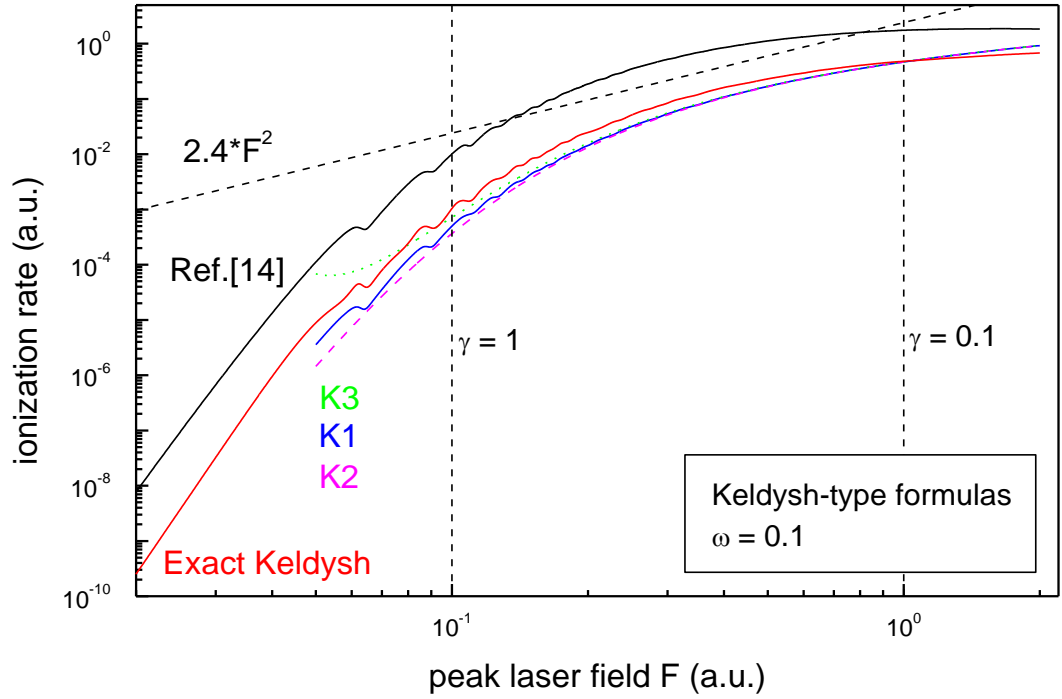

FIG. 8. (Color online) Total ionization rates as a function of the peak laser field for $\omega = 0.1$ *a.u.*. Dashed black (slanted) line: numerical fit obtained in Ref. [22]. Solid black line: the theory from Ref. [14]. Original Keldysh theories are denoted (from top to bottom in the plot) as "K3" (dotted green line), "K1" (solid blue line), and "K2" (dashed magenta line). Solid red line: "Exact Keldysh" theory. Two vertical dashed lines show values of the Keldysh parameter (for a given $F$), which decreases from left to right.

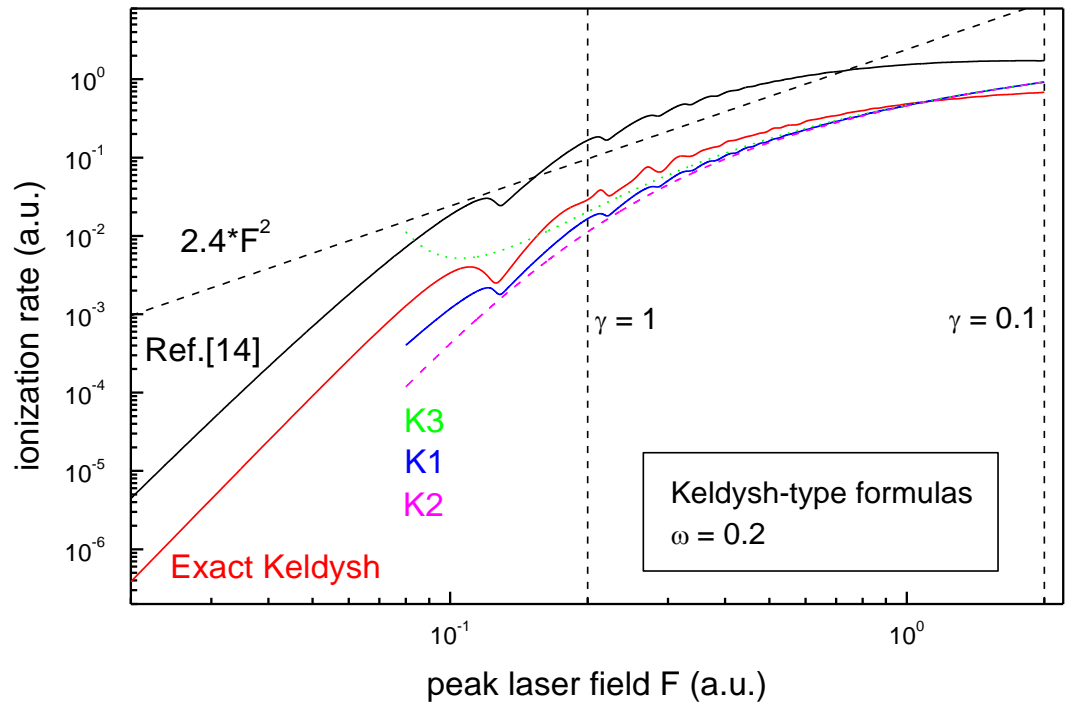

FIG. 9. (Color online) Same as Fig. 8, but for $\omega = 0.2$ *a.u.*.